\documentclass[aps,pra,reprint,showpacs,superscriptaddress]{revtex4-1}
\usepackage[english]{babel}
\usepackage{amsmath,amssymb,bbm,graphicx,color,comment,txfonts}
\usepackage[bookmarks=true,colorlinks,citecolor=blue,urlcolor=blue]{hyperref}
\usepackage{dsfont}

\newcommand{\dph}{\gamma_{\text{de}}}
\newcommand{\dg}{\gamma_{\downarrow g}}
\newcommand{\dr}{\gamma_{\downarrow 0}}

\newcommand{\eq}[2]{\begin{eqnarray}\label{#1} #2 \end{eqnarray}}

\begin{document}

\date{\today}

\title{Controlling excitation avalanches in driven Rydberg gases}
\author{Kai Klocke}
\affiliation{Department of Physics and Institute for Quantum Information and Matter, California Institute of Technology, Pasadena, CA 91125, USA}
\author{Michael Buchhold}
\affiliation{Department of Physics and Institute for Quantum Information and Matter, California Institute of Technology, Pasadena, CA 91125, USA}

\begin{abstract}
Recent experiments with strongly interacting, driven Rydberg ensembles have introduced a promising setup for the study of self-organized criticality (SOC) in cold atom systems. Based on this setup, we theoretically propose a control mechanism for the paradigmatic avalanche dynamics of SOC in the form of a time-dependent drive amplitude. This gives access to a variety of avalanche dominated, self-organization scenarios, prominently including self-organized criticality, as well as sub- and supercritical dynamics. 
We analyze the dependence of the dynamics on external scales and spatial dimensionality. It demonstrates the potential of driven Rydberg systems as a playground for the exploration of an extended SOC phenomenology and their relation to other common scenarios of SOC, such as e.g., in neural networks and on graphs.
\end{abstract}

\maketitle
\section{Introduction} Away from thermal equilibrium and in the absence of detailed balance, (quasi-) stationary states emerge from ordering principles different from the equipartition of energy. Outstanding amongst such out-of-equilibrium ordering mechanism is self-organized criticality. Introduced in the seminal paper of Bak, Tang and Wiesenfeld (BTW)~\cite{Bak1987,Bak1988} to explain the emergence of flicker noise in electrical circuits, SOC has since then been observed in a variety of diverse, mainly large scale systems, ranging from earth quakes~\cite{Sornette1989,Bak1991,Bak2002}, forest fires~~\cite{Schenk2000,Drossel1992,Malamud1998,Turcotte1999} and solar flares~\cite{Lu1991,Aschwanden2016} to vortex dynamics in superconductors~\cite{Field1995,Altshuler2004} and turbulence~\cite{Chapman2009}. Only recently, SOC was recognized as a possible mechanism to establish optimal conditions for information spreading~\cite{Arca2006,Shew2015,Hesse2014,Markovic2014,Kinouchi2006,Levina2007}. 

The phenomenon of SOC can be described by simple means, by balancing dissipation and external drive, a many-body system is attracted, i.e., it self-organizes, towards a state with scale invariant correlations~\cite{Lu1995,Watkins2016,dickman2000}. In thermal equilibrium, scale invariance is associated with dynamics at a critical point signaling a continuous phase transition~\cite{ZinnJustinBook}. Compared to a fine tuned critical point, scale invariance due to SOC is believed to occur in an extended parameter regime, commonly enabled by a separation of time scales between drive and dissipation~\cite{Vespignani1997,dickman2000,Watkins2016}. While this makes SOC robust to changes in the external conditions, the interplay of interactions, drive and dissipation obscure its origin and only few microscopic models are found in the literature.

Apart from the sandpile model of BTW, manifestations of SOC in nature are mostly approached via phenomenological models~\cite{Malamud1998,Rybarsch2014}, either because the microscopic description is too complex or the elementary building blocks are unknown~\cite{Rhodes1996,Shew2015,Aschwanden2018}. Unfortunately, many realizations of SOC don't match the  energy conserving dynamics of BTW's sandpile model. This makes both the microscopic understanding and, even more, the controllability of SOC extremely challenging~\cite{Watkins2016,dickman2000}. This applies especially to neural network dynamics, which are entirely based on effective models~\cite{Strogatz,Barzel2013}. Consequently, a setup exploring an extended SOC phenomenology on the one hand and featuring the knowledge and a large degree of controllability of its basic elements on the other hand represents a promising tool to study aspects of SOC in generic nonequilibrium settings.
\begin{figure}\includegraphics[width=\linewidth]{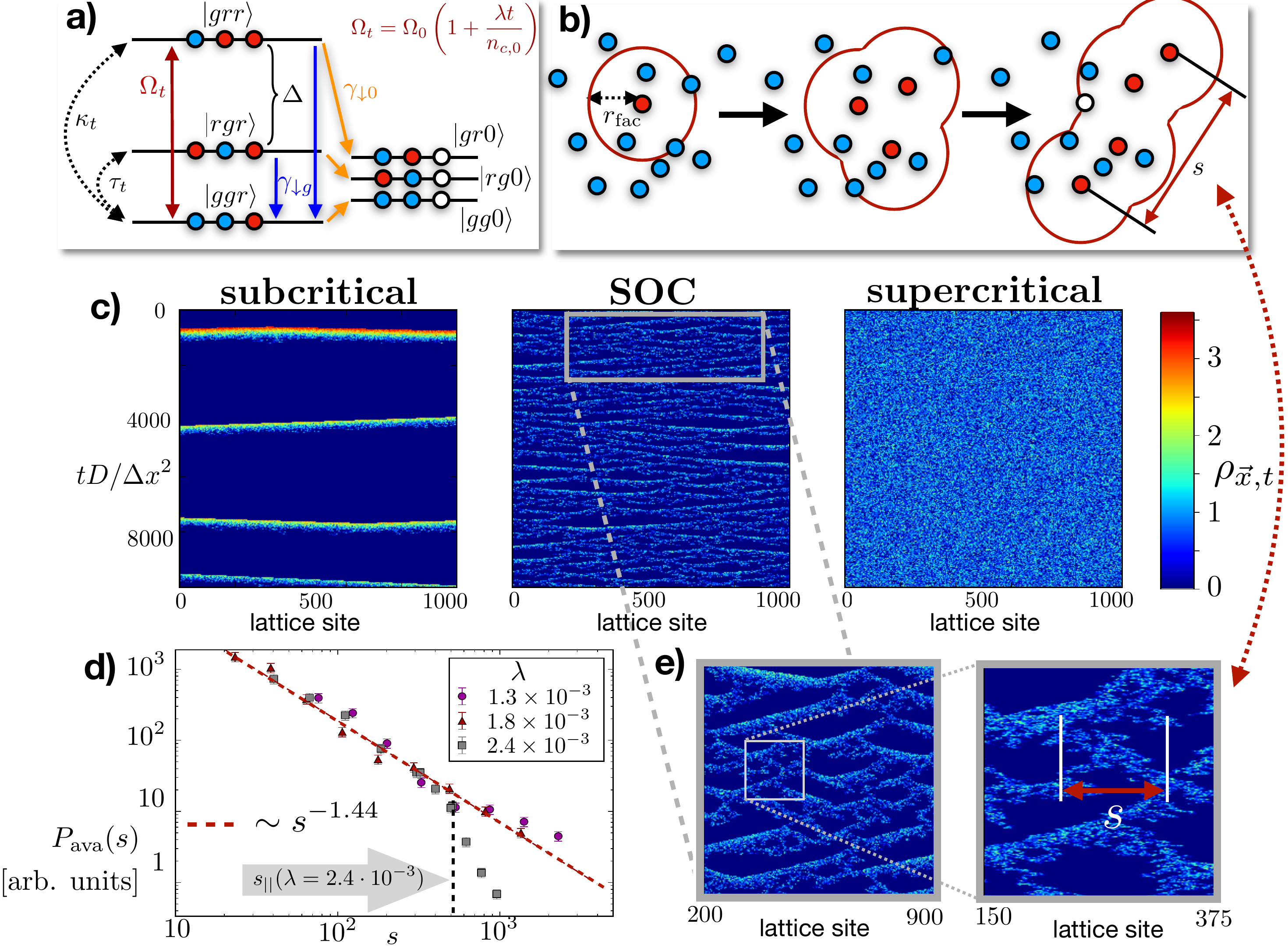}\caption{Driven Rydberg self-organized criticality. (a) Three-atom level scheme: transitions from the ground $|g\rangle$ to the Rydberg state $|r\rangle$ are only resonant inside the facilitation radius $r_{\text{fac}}$ of a second Rydberg atom. (b) Illustration of an excitation avalanche triggered by a single Rydberg atom (red dots) in two dimensions. After a period $t\sim \kappa_t^{-1}$, Rydberg atoms facilitate the excitation of ground state atoms (blue dots) inside the facilitation radius (red line), creating avalanches of length $s$ before decaying into the ground or removed state (white dot) with rates $\Gamma, \dr$.  (c)  Real space dynamics of the Rydberg density $\rho_{\vec{x},t}$ on a one-dimensional grid of $N=10^3$ sites ($x$-axis) with time progressing along the $y$-direction. Depending on the pumping strength growth rate $\lambda$, avalanches form a periodic (subcritical) structure, a fractal (SOC) structure or a random noise pattern (supercritical). (d) Distribution of avalanche sizes $s$ (logarithmic scale) in the SOC regime (dots and triangles) and at the transition to the supercritical regime (squares). (e) Zoom-in to the SOC pattern and illustration of the length $s$ of an individual avalanche in the one-dimensional setting. We define $s$ as the total length of an isolated avalanche before it depletes or merges with other avalanches.}\label{Fig1}\end{figure}

Only recently a promising candidate has been introduced in an experiment with a gas of driven Rydberg atoms \cite{Buchhold2018a}; above a certain driving threshold, the atomic pseudo-spins self-organize towards a transient, scale invariant state, featuring common signatures of SOC~\cite{dickman2000,Aschwanden2018}. Our work builds up on this basic setting for SOC in cold Rydberg gases. 

We propose the implementation of a control mechanism for excitation avalanches in driven Rydberg ensembles and explore the corresponding many-body dynamics. We show how this gives access to an extended SOC phenomenology, including subcritical and supercritical avalanche dynamics. By adjusting the proposed mechanism to common control parameters such as the laser intensity and the detuning, one can access the paradigms of SOC: a scale invariant avalanche distribution~\cite{Turcotte1999} and a $\frac{1}{\omega}$-noise pattern~\cite{Bak1987,Bak1988}. 

\section{Facilitated Rydberg dynamics} We consider the many-body dynamics in a gas of interacting Rydberg atoms~\cite{schaust,Helmrich2013,Gorshkov2013,Helmrich2018,Letscher2017,Thomas2018}, which move freely inside a trap. Each Rydberg atom is modeled as an effectively three level system, consisting of a non-interacting ground state $|g\rangle$, a highly polarizable Rydberg state $|r\rangle$ with large principle quantum number $n\gg1$~\cite{Gallagher84,RevModPhys.82.2313,Weimer2012} and an auxiliary, removed state $|0\rangle$. The latter is a container state representing a set of internal states that can be reached via dissipative decay but are otherwise decoupled from the $|g\rangle-|r\rangle$ sector~\cite{Helmrich2018,Buchhold2018a}. Each atom obtains a label $l$ and a set of operators $\sigma^{ab}_l\equiv |a\rangle\langle b|_l$ acting on its internal states.

The ensemble is subject to a laser, coherently driving the $|g\rangle - |r\rangle$ transition with a Rabi frequency $\Omega$ and detuning from resonance $\Delta$. The highly excited Rydberg state is subject to dissipation originating from dephasing as well as spontaneous decay into both the ground state $|g\rangle$ and the removed state manifold $|0\rangle$ with effective rates labeled by $\dph, \dg, \dr$~\cite{Buchhold2018a}. Due to their polarizability, two atoms, labeled $l,l'$, in the Rydberg state experience a mutual van-der-Waals repulsion. Its potential form is $V^I_{l,l'}=C_6 |\vec{r}_l-\vec{r}_{l'}|^{-6}$, where $C_6$ is the van-der-Waals coefficient and $\vec{r}_l, \vec{r}_{l'}$ are the atomic positions \cite{pfau2013}\footnote{The interaction might as well acquire a dipole-dipole form, $V\sim|\vec{r}_l-\vec{r}_{l'}|^{-3}$, e.g., due to Foerster resonances~\cite{Li2005}. This does, however, not modify the structure of Eq.~\eqref{Eq4}.}. 

As a simple but crucial innovation we consider here a time-dependent Rabi frequency \eq{Extra1}{\Omega\rightarrow\Omega_t=\Omega_0(1+t\frac{\lambda}{2 n_{c,0}}),} with an initial frequency $\Omega_0$, a dimensionless density $n_{c,0}$, which we define later, and a ramp parameter $\lambda\ll n_{c,0}\Omega_0$. This corresponds to a slow, linear increase of the pump laser intensity $I_t\sim \Omega_0^2(1+\frac{\lambda t}{n_{c,0}}+O(\lambda^2t^2)$. It gives rise to a continuously increasing excitation probability for the $|g\rangle \leftrightarrow|r\rangle$ transition, counteracting the decay into the removed state and balancing the system at a fixed, non-zero density of excited states for transient times $t<\frac{n_{c,0}}{\lambda}$. 

The microscopic dynamics of the $d$-dimensional gas are given by the master equation  ($\hbar=1$) \eq{Extra2}{\partial_t\hat{\rho}=i [\hat\rho,H]+\sum_l \mathcal{L}_l\hat\rho} for the ensemble density matrix $\hat{\rho}$. The coherent atom-light and atom-atom interaction is captured by the Hamiltonian
\begin{eqnarray}\label{Eq2}H=\sum_l \left\{\left(\sum_{l'\neq l}\frac{C_6}{2|\vec{r}_l-\vec{r}_{l'}|^6}\sigma^{rr}_{l'}-\Delta \right)\sigma^{rr}_l+\frac{\Omega_t}{2}\left(\sigma^{rg}_l+\sigma^{gr}_l\right)\right\}, \ \ \end{eqnarray}
while dissipative processes are described by the Liouvillian 
\begin{eqnarray}\label{Eq3}\mathcal{L}_l\hat\rho&=&\dph \sigma^{rr}_l\hat\rho\sigma^{rr}_l+\dg\sigma^{gr}_l\hat\rho\sigma^{rg}_l+\dr\sigma^{0r}_l\hat\rho\sigma^{r0}_l-\frac{\Gamma}{2}\{\sigma^{rr}_l,\hat\rho\},\ \ \ \ \ \end{eqnarray}
where $\Gamma=\dph+\dg+\dr$ is the sum of all dissipative rates. In typical experiments~\cite{Helmrich2018,Buchhold2018a,Morsch2016}, the product of the atomic mass $M$ and temperature $T$ is 'large' compared to the density $n_0$, causing a thermal de-Broglie wavelength $\lambda_{\text{th}}=\frac{h}{\sqrt{2\pi M k_{\text{B}}T}}$ much smaller than the mean free path $d_a\sim n_0^{-1/d}$. The motional degrees of freedom $\vec{r}_{l,l'}$ thus cannot maintain coherence between two subsequent scattering events and are treated as classical variables undergoing thermal motion, see below and Ref.~\cite{Buchhold2018a}.

We focus on a very large detuning $\Delta/\Gamma \sim O(10^2-10^3)$~\cite{Urvoy2015a,Gart2013,TonyLee2012}, leading to strongly suppressed, off-resonant single particle transitions $|g\rangle \leftrightarrow |r\rangle$ at a rate $\tau_t\equiv\frac{\Gamma\Omega_t^2}{\Gamma^2+4\Delta^2}$. Due to interactions, an atom in the Rydberg state, however, creates a facilitation shell of radius $r_{\text{fac}}=(C_6/\Delta)^{1/6}$ and width $\delta r_{\text{fac}}\sim r_{\text{fac}}\Gamma/\Delta$. Inside the shell, the Rydberg repulsion compensates the detuning in Eq.~\eqref{Eq2}, yielding an effective resonant excitation rate $\kappa_t\approx \Omega_t^2/\Gamma$ with $\kappa_t\gg\tau_t$~\cite{Ates2007a,Amthor2010,Lesa2014,Morsch2016a,Marcuzzi2016}. 

In the limit of strong dephasing, the atom coherences decay rapidly in time and the relevant dynamical degrees of freedom are the Rydberg state density and the density of 'active' states, i.e., of atoms in the Rydberg {\it and} in the ground state. Their coarse grained values, averaged over a 'facilitation cluster' of volume $V_{\text{fac}}= \frac{\pi^{\frac{d}{2}}r_{\text{fac}}^d}{\Gamma_{\text{Euler}}(\frac{d}{2}-1)}$ are 
\eq{Extra3}{\rho_{\vec{x},t}&\equiv&\sum_{l\text{ s.t. } |\vec{r}_l-\vec{x}|\le r_{\text{fac}}} \langle \sigma^{rr}_l\rangle(t), \\ \label{Extra3b}n_{\vec{x},t}&\equiv&\sum_{l\text{ s.t. } |\vec{r}_l-\vec{x}|\le r_{\text{fac}}} \langle \sigma^{rr}_l+\sigma^{gg}_l\rangle(t).} 
The evolution equations for $\rho_{\vec{x},t}$ and $n_{\vec{x},t}$ are obtained by adiabatically eliminating the atom coherences from the Heisenberg equations of motion \cite{Marcuzzi2015,Marcuzzi2016,Morsch2016,Buchhold2017,PerezEspigares2017,Morsch2017}. This yields the Langevin equation (see ~\cite{Buchhold2017,Buchhold2018a})
\begin{eqnarray}\label{Eq4}\partial_t\rho_{\vec{x},t}=D\nabla^2\rho_{\vec{x},t}+(\kappa_t\rho_{\vec{x},t}+\tau_t)(n_{\vec{x},t}-2\rho_{\vec{x},t})-\Gamma\rho_{\vec{x,t}}+\xi_{\vec{x},t}. \ \ \ \end{eqnarray}

{\color{black}Equation \eqref{Eq4}, describes four different processes on a coarse grained time scale $t\sim \dph^{-1}$. It covers the average over Rabi oscillations inside each cluster, which occur with rate $\kappa_t\rho_{\vec{x},t}+\tau_t$ and prefer (averaged over time) a semi-excited state $\rho_{\vec{x},t}=\frac{n_{\vec{x},t}}{2}$. The rate combines the off-resonant oscillation rate $\tau_t$ and the resonant, facilitated rate $\kappa_t\rho_{\vec{x},t}$, which is proportional to the number of facilitating atoms $\rho_{\vec{x},t}$. This process competes with the linear decay channel $\sim\Gamma$, which prefers the ground state $\rho_{\vec{x},t}=0$.}

The spreading of excitations from cluster to cluster is described by the diffusion term $\sim D\nabla^2\rho_{\vec{x},t}$, with $D=\kappa_t S$ being proportional to the facilitation rate and the surface $S$ of the clusters \cite{Buchhold2018a}. In a dissipative environment, each cluster experiences fluctuations of $\rho_{\vec{x},t}$, which are proportional to the oscillation rate \cite{Marcuzzi2015,Marcuzzi2016,Morsch2016,Buchhold2017} and covered by the Markovian noise kernel (overline indicating noise average) \eq{Extra4}{\overline{ \xi_{\vec{x},t}\xi_{\vec{y},t'}}=\delta(\vec{x}-\vec{y})\delta(t-t')\left[(\tau+\kappa \rho_{\vec{x},t})n_{\vec{x},t}+2\Gamma\rho_{\vec{x},t}\right].} 

Before turning to the evolution of $n_{\vec{x},t}$, we discuss the mean-field solution of Eq.~\eqref{Eq4} in the limit where $\tau_t\ll \Gamma,\kappa_tn_{\vec{x},t}$ by setting $D=\xi_{\vec{x},t}=0$. Defining a {\it critical} density $n_{c,t}\equiv\frac{\Gamma}{\kappa_t}$, one distinguishes two different regimes: an {\it inactive} regime for $n_{\vec{x},t}<n_c$, where the Rydberg density is suppressed and evolves towards $\rho_{\vec{x},t}\rightarrow \frac{\tau}{\Gamma}n_{\vec{x},t}$, and an {\it active} regime for $n_{\vec{x},t}>n_c$, where it evolves towards $\rho_{\vec{x},t}\rightarrow \frac{1}{2}(n_{\vec{x},t}-n_c)$. The crossover between the two regimes at $n_{\vec{x},t}=n_{c,t}$ features a maximal correlation length of $\xi_{||}=\sqrt{\frac{D}{\sqrt{8\Gamma\tau_t}}}$. It turns into a sharp, second order phase transition in the limit $\tau_t\rightarrow0$ \cite{Janssen1981,Hinrichsen2000,Marcuzzi2016,Marcuzzi2015,Buchhold2017}. 

{\color{black} The above discussed mean-field solution illustrates the dynamics in the regimes $\Gamma\ll \kappa_t n_{\vec{x},t}$ and $\Gamma\gg \kappa_t n_{\vec{x}t}$. In the presence of spatial fluctuations, i.e., for $D>0$, the asymptotic values for $\rho_{\vec{x},t}$ and $n_{\vec{x},t}$ above remain good approximations far away from the critical point $|\kappa_t n_{\vec{x},t}-\Gamma|\gg 1$. For $n_{\vec{x},t}\rightarrow n_{c,t}$, however, spatial fluctuations, manifesting via propagating avalanches with strongly fluctuating density, become increasingly strong and lead to deviations of the uniform behavior. In addition, the critical density is generally shifted towards larger values $n_c>\Gamma/\kappa$. In order to determine $n_c$ in $d=1,2$, we compute the location of the critical point in Eq.~\eqref{Eq4} numerically, e.g., we find $n_c=3.86$ in $d=1$.}

The evolution of the density $n_{\vec{x},t}$ is governed by thermal motion of the atoms, the decay into the removed state and density fluctuations. It is summarized in the Langevin equation ~\cite{Buchhold2018a}
\begin{eqnarray}\label{Eq5}\partial_t n_{\vec{x},t}=D_n \nabla^2 n_{\vec{x},t}-\dr \rho_{\vec{x},t}+\eta_{\vec{x},t}\end{eqnarray} with a Markovian noise kernel $\langle\eta_{\vec{x},t}\eta_{\vec{y},t'}\rangle=\delta(\vec{x}-\vec{y})\delta(t-t')\dr\rho_{\vec{x},t}$ and a thermal diffusion constant $D_n$. It has minor impact on the dynamics but reduces geometrical constraints due to rare, inhomogeneous configurations of $n_{\vec{x},t}$ \footnote{Any rare configuration with $n_{\vec{x},t}=0$ would otherwise block the spreading of excitations forever.}. 

\section{Derivation of the Langevin equations} In this section, we present the detailed derivation of the Langevin equations \eqref{Eq4} and \eqref{Eq5} from the master equation \eqref{Extra2}. Readers interested in the effective dynamics may continue with its discussion in the following section. 

Due to the exponential growth of the Hilbert space, the master equation Eq.~\eqref{Extra2} becomes too complex to solve for realistic, macroscopic system sizes. In order to reduce the complexity, the dynamics are projected onto the relevant long-wavelength degrees of freedom, i.e., the Rydberg density $\rho$ and the active density $n$ as defined in Eqs.~\eqref{Extra3} and \eqref{Extra3b}. This procedure has been discussed for the case of $\lambda=\dr=0$ in Refs.~\cite{Marcuzzi2016, Buchhold2017} and for the case $\lambda=0, \dr\neq0$ in Ref.~\cite{Buchhold2018a}. 

For strong dephasing $\dph\gg \Omega_t$ the decay of the atomic coherences $\sigma_{l}^{rg}, \sigma_{l}^{r0}$  towards their steady state value is the fastest process in the quantum master equation. They can be adiabatically eliminated by formally solving the steady state equation for the average ($\alpha=g,0$)
\eq{AEq1}{0\overset{!}{=}\partial_t \langle \sigma_{l}^{r\alpha}\rangle=\text{Tr}\left[\sigma_{l}^{r\alpha}\left(i[\hat{\rho},H]+\sum_l\mathcal{L}_l\hat{\rho}\right)\right].} 
Inserting the solution of Eq.~\eqref{AEq1} and the completeness relation $\sigma_{l}^{rr}+\sigma_{l}^{gg}+\sigma_{l}^{00}=\mathds{1}$ into the full Heisenberg-Langevin equations for $\sigma^{rr}_l, \sigma^{gg}_l$ yields
\eq{AEq3}{
\partial_t \sigma^{gg}_l&=&-\partial_t\sigma^{rr}_l-\dr \sigma^{rr}_l+\xi^g_l,\\
\partial_t\sigma^{rr}_l&=&\frac{\Omega_t ^2 \Gamma(\sigma_{l}^{gg}-\sigma_{l}^{rr})}{\Gamma^2+4(\Delta-\sum_{l'\neq l} V^I_{l,l'}\sigma^{rr}_{l'})^2}-\Gamma  \sigma^{rr}_l+\xi_l^r.\ \ \ \ \ \ \label{AEq4}
}
The Markovian noise operators $\xi_l^{r,g}$ are added in order to enforce the fluctuation-dissipation relation of the driven dissipative master equation. They are local in space and time and fulfill the generalized Einstein relation {\color{black}(overline indicating noise average)
\eq{Eq8}{
\overline{(\xi_l^{r})^2}&=&\underbrace{\overline{\partial_t(\sigma_{l}^{rr})^2}}_{=\partial_t\overline{\sigma^{rr}_l} \text{ since } (\sigma^{rr}_l)^2=\sigma^{rr}_l}-2\overline{\sigma_{l}^{rr}  \partial_t \sigma_{l}^{rr}} .
}
This noise average leads to the $\delta$-correlated Markovian noise in Eq.~\eqref{Extra4} for the Langevin equation after the coarse graining procedure. Its crucial property for the realization of SOC is the scaling of the noise $\sim \rho_{\vec{x},t}$ (except for the tiny fluctuations $\sim \tau_t n_{\vec{x},t}$), which is responsible for a well defined, fluctuationless inactive phase. }

Since the operators $\sigma^{rr}_l, \sigma^{gg}_l$ are projection operators with eigenvalues $0,1$, any function $f$ of, say $\sigma^{rr}_l$, can be expressed as $f(\sigma^{rr}_l)=f(0)+(f(1)-f(0))\sigma^{rr}_l$. Extending this to the whole set of $\{\sigma^{rr}_l, \sigma^{gg}_l\}$, one rewrites 
\eq{AEq5}{
&&\frac{\Omega_t ^2 \Gamma}{\Gamma^2+4(\Delta-\sum_{l'\neq l} V^I_{l,l'}\sigma^{rr}_{l'})^2}=\underbrace{\frac{\Omega_t ^2 \Gamma}{\Gamma^2+4\Delta^2}}_{
=\tau_t}+ \nonumber\\ 
&&\sum_{l'\neq l}\left(\frac{\Omega_t ^2 \Gamma}{\Gamma^2+4(\Delta-V^I_{l,l'})^2}-\tau_t\right)\sigma_{l'}^{rr}+O(\sigma^{rr}_{l'}\sigma^{rr}_{l''}).\label{AEq6}
}
This expression is 
exact up to second order powers in the projection operators. It separates off-resonant single particle transitions with rate $\tau$ and facilitated, two-particle transitions. For $2|\Delta-V^I_{l,l'}|<\Gamma$, the facilitation rate deviates significantly from zero. Depending on the interaction potential, this defines the facilitation radius $r_{\text{fac}}$, i.e., for a typical van der Waals potential $V_{l,l'}=\frac{C_6}{r^6}$ one finds $r_{\text{fac}}\equiv(\rm{C}_6/\Delta)^{1/6} $ and the facilitation shell $|\vec{r}_l-\vec{r}_{l'}|\in [r_{\text{fac}}-\Delta r_{\text{fac}}, r_{\text{fac}}+\Delta r_{\text{fac}}]$ with $\Delta r_{\text{fac}}=r_{\text{fac}}\frac{\Gamma}{12\Delta}$. We introduce a real space projector $\Pi_{ll'}$ with $\Pi_{ll'}=1$ if $|\vec{r}_l-\vec{r}_{l'}|$ is inside the facilitation shell and zero otherwise. This yields
 \eq{Eq6}{
\partial_t\sigma^{rr}_l&=&\left(\tau+\frac{\Omega_t^2}{\Gamma}\sum_{l'\neq l}\Pi_{ll'}\sigma^{rr}_{l'}\right)(\sigma^{gg}_l-\sigma^{rr}_l)-\Gamma  \sigma^{rr}_l+\xi^r_l.\ \ \ }

This provides a good approximation for the facilitation rate when the density of excitations is small. {\color{black} For a number of $m\ge1$ excited states inside a single shell, however, the exact solution shows a growth of the shell radius as {\color{black}$r_{\text{fac}}^{(m)}=m^{1/6}r_{\text{fac}}$} (in $d=3$ dimensions). This scaling behavior could be either taken into account by expanding Eq.~\eqref{AEq6} up to higher orders in the $\sigma^{rr}$ operators, which would account for a larger number $m>1$ of excitations per cluster, or by including the scaling of the facilitation volume for $m>1$ particles compared to the case of  $m=1$. In both cases, the facilitation rate for $m>1$ then grows $\propto\sqrt{m}$, compared to the $\propto m$ prediction of Eqs.~\eqref{Eq6} and \eqref{AEq6}.If one bears in mind, however, the weak off-resonant excitation rate, configurations of $m\ge1$ are suppressed by a factor $o(10^{-4})$. Our simulations show that $\rho_{\vec{x},t}\le1$ in most cases, which validates the restriction to $m=0,1$ in Eqs.~\eqref{Eq6} and \eqref{AEq6}.}

The equation of motion for $\rho_{\vec{x},t}=\sum_{l}\Theta(r_{\text{fac}}-|\vec{x}-\vec{r_l}|)\langle\sigma^{rr}_l\rangle$ and $n_{\vec{x},t}=\sum_{l}\Theta(r_{\text{fac}}-|\vec{x}-\vec{r_l}|)\langle\sigma^{rr}_l+\sigma^{gg}_l\rangle$ yields
\eq{Eq7}{
\partial_t\rho_{\vec{x},t}=\sum_{l}\left(\langle\partial_t\sigma^{rr}_l\rangle+\langle\sigma^{rr}_l\rangle\partial_t\vec{r}_l\vec{\nabla}\right)\Theta(r_{\text{fac}}-|\vec{x}-\vec{r_l}|)
}
and similar for $n_{\vec{x},t}$. For a homogeneous density, the drift term $\sim \partial_t\vec{r}_l$ can be approximated to be zero (see below for an inhomogeneous setting). This yields
\eq{Eq8}{
\partial_t n_{\vec{x},t}&=&-\dr\rho_{\vec{x},t}+\eta_{\vec{x},t},\nonumber\\
\partial_t\rho_{\vec{x},t}&=&\left(\tau_t+\frac{\Omega_t^2}{\Gamma}\mathcal{F}_{\vec{x}}(\rho_{\vec{z},t})\right)(n_{\vec{x},t}-2\rho_{\vec{x},t})-\Gamma\rho_{\vec{x},t}+\xi_{\vec{x},t},
}
where $\mathcal{F}_{\vec{x}}(\rho_{\vec{z}},t)$ is some linear, quasi-local functional of $\rho_{\vec{x},t}$.

 $\mathcal{F}_{\vec{x}}(\rho_{\vec{z},t})$ has support only around $|\vec{x}-\vec{z}|=r_{\text{fac}}$, enabling a Taylor expansion of the density. Since the Rydberg facilitation mechanism is isotropic in space, the expansion contains only even powers of derivatives. It reads as [cf. Eq.~(14) in Ref.~\cite{Buchhold2018a}]
\eq{Eq15}{
\mathcal{F}_{\vec{x}}(\rho_{\vec{z},t})=\mathcal{F}_{\vec{x}}(1)\rho_{\vec{x},t}+\frac{\mathcal{F}_{\vec{x}}(\vec{z}^2)}{2}\nabla^2\rho_{\vec{x},t}+O(\nabla^4\rho_{\vec{x},t}).
}

 The noise $\langle \xi_{\vec{x},t}\xi_{\vec{y},s}\rangle=\sum_{l,m}\Theta(r_{\text{fac}}-|\vec{x}-\vec{r_l}|)\Theta(r_{\text{fac}}-|\vec{y}-\vec{r_m}|)\langle\xi_{l,t}\xi_{m,s}\rangle
=\delta(s-t)\delta(|\vec{x}-\vec{y}|)\left[\kappa_t \rho_{\vec{x},t}+\tau_t\right]$
remains Markovian and $\delta$-correlated on length scales of the facilitation radius.

Making a conservative estimate for the temperature of the motional degrees of freedom $T=O(10\mu K)$ and the atomic mass $M=O(20$u)~\cite{Buchhold2018a}, one finds a thermal de Broglie wavelength $\lambda_T=\frac{h}{\sqrt{2\pi Mk_\text{B}T}}\approx200$nm. For an atomic density of $n_0\approx 10^{11}\text{cm}^{-3}$ the mean free path in three dimensions amounts to $d_a=\Big(\frac{6}{\pi n_0}\Big)^{1/3}\sim 2\mu$m, which is at least one order of magnitude larger than $\lambda_T$. Consequently, coherence in the motional degrees of freedom is lost between two subsequent scattering events and they can be treated classically. In the absence of an external trapping potential, the particles perform Brownian motion, i.e., thermal diffusion in a dilute van der Waals gas. This allows us to treat the atomic positions as slowly diffusing and uniformly distributed in space. 

Including Brownian motion with diffusion constant $D_n$ the final form of the Langevin equations is
\eq{Eq17}{
\partial_t n_{\vec{x},t}&=&D_n\nabla^2n_{\vec{x},t}-\dr\rho_{\vec{x},t}+\xi_{\vec{x},t},\nonumber\\
\partial_t\rho_{\vec{x},t}&=&D\nabla^2\rho_{\vec{x},t}+(\kappa_t\rho_{\vec{x},t}+\tau_t)(n_{\vec{x},t}-2\rho_{\vec{x},t})-\Gamma\rho_{\vec{x},t}+\xi_{\vec{x},t}.\ \ \ 
}
Here $\kappa_t=\mathcal{F}_{\vec{x}}(1)\frac{\Omega_t^2}{\Gamma}$ is the facilitation rate. The diffusion constant   $D=\mathcal{F}_{\vec{x}}(\vec{z}^2)\frac{\Omega_t^2}{2\Gamma}(n_{\vec{x},t}-2\rho_{\vec{x},t})+D_n\approx\mathcal{F}_{\vec{x}}(\vec{z}^2)\frac{\Omega_t^2}{2\kappa_t}$ is dominated by the facilitated spreading, which is proportional to the average density, i.e., $n_{\vec{x},t}-2\rho_{\vec{x},t}\approx \frac{\Gamma}{\kappa_t}$. This makes $D$, apart from local density fluctuations, time independent.

\section{Self-organized criticality and avalanche dynamics} In order to observe self-organization towards a long-range correlated state, the dynamics should push any initial density $n_{\vec{x},0}$ close towards $n_{\vec{x},t}\rightarrow n_{c,t}$ and thereby maximize the correlation length $\xi_{||}$. This is achieved by the combination of loss into the auxiliary state $\sim \dr$ and the continuously growing pump strength $\sim \lambda$. 

Their interplay is best understood by expanding the critical density $n_{c,t}$ up to first order in $\lambda t$, yielding 
\eq{TaylorCrit}{
n_{c,t}=\frac{\Gamma}{\kappa_t}=n_{c,0}-\left. \frac{\partial\kappa_t}{\partial t}\right|_{t=0}\frac{t\cdot n_{c,t}}{\kappa_t}=n_{c,0}-\lambda t,
} which is valid for $\lambda t<n_{c,0}$. For active densities $n_{\vec{x},t}\approx n_{c,t}$, the Rydberg state density $\rho_{\vec{x},t}$ experiences a large correlation length, leading to long-lived and and far spreading excitations, i.e., the formation of avalanches. Once an avalanche has formed, parts of it decay into the removed state, leading to a decrease of $n_{\vec{x},t}$. It reaches a stationary point when the decay of both $n_{\vec{x},t}$ and $n_{c,t}$ compensate each other, i.e., for $\lambda=\dr\rho_{\vec{x},t}$.

On times $t<\frac{\lambda}{n_{c,0}}$, this is the only homogeneous solution of Eqs.~\eqref{Eq4} and \eqref{Eq5} with \eq{Extra5}{\rho_{\vec{x},t}=\frac{\lambda}{\dr} \text{ and } n_{\vec{x},t}=n_{c,t}+\frac{2\lambda}{\dr}+\frac{\dr\tau_t}{\kappa\lambda n_{c,t}}.} It is reached after a time $t\approx \max\{\kappa_t^{-1},\dr^{-1}\}$ and it survives up to times of order $t\approx \frac{n_{c,0}}{\lambda}$. On larger times, effects of order $\lambda^2t^2$ set in and the active density depletes to zero, i.e., $\rho_{\vec{x},t},n_{\vec{x},t}\rightarrow0$.

Imposing a double separation of time scales on the dynamics via 
\eq{Conditions}{
\text{\it (i) } \frac{\tau_t}{\lambda}\rightarrow 0^+ \text{ and {\it (ii)} } \frac{\lambda}{\dr}\rightarrow0^+,}  Eq.~\eqref{Extra5} predicts the self-organization towards a long-lived and long-range correlated state with $\rho_{\vec{x},t}= 0^+$,  $n_{\vec{x},t}= n_{c,t}+0^+$ and $\xi_{||}\rightarrow\infty$. We thus call {\it (i) + (ii)} the conditions for SOC in our driven Rydberg setup. The degree up to which both conditions are met, i.e., SOC is realized, can be adjusted experimentally via the Rabi frequency $\Omega_t$, the detuning $\Delta$ or the decay $\dr$. 

\begin{figure}[t]\includegraphics[width=\linewidth]{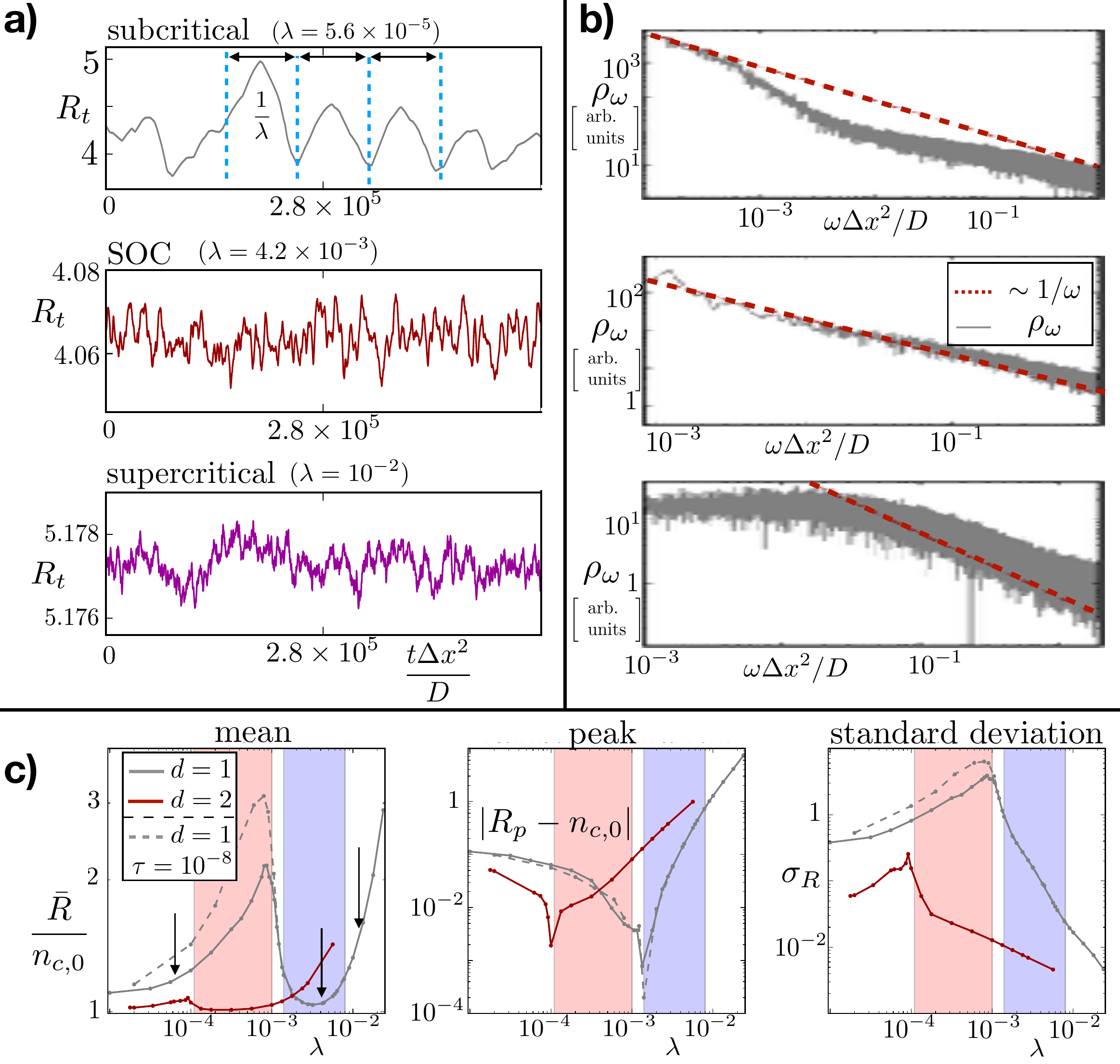}\caption{Experimental observables. (a) Time evolution of the integrated density $R_t$, Eq.~\eqref{Extra7}, in three different regimes ($n_{c,0}\approx 4$ for comparison).(b) Fourier decomposition $\rho_\omega$ of the Rydberg density, same parameters as in (a). (c) Time averaged mean $\bar R$, standard deviation $\sigma_R$ and peak value of the integrated density $R_t$ in dimensions $d=1,2$. A sharp drop of $\bar R, \sigma_R$ marks the onset of SOC, i.e., a regime of scale invariant avalanche distributions (colored region, with blue corresponding to $d=1$ and red to $d=2$). Arrows indicate the values of $\lambda$ used in the plots (a) and(b).}\label{Fig2}
\end{figure}

Such double separation of scales is a common requirement for realizations of SOC without energy conservation \cite{Bonachela2009,Bonachela2010} \footnote{This is contrasted with SOC in energy conserving systems, e.g., the sandpile model, which requires only a single pair of separated scales \cite{Bonachela2009}.}. Since both our Hamiltonian and the Lindblad dynamics do not conserve the energy, the conditions {\it (i)+(ii)} can be seen as the present manifestations of this phenomenon. One may now argue that such strict requirements do not really differ from parameter fine tuning in conventional criticality. We, however, show that the dynamics of Eqs.~\eqref{Eq4} and \eqref{Eq5} display SOC even for very weak realizations of {\it (i)} and {\it (ii)}, e.g., for $\frac{\tau_t}{\lambda}\sim10^{-4}$ and $\frac{\lambda}{\dr}\sim0.1$, making it accessible to experiments.

We emphasize that for $t<\frac{n_{c,0}}{\lambda}$ the increase of $\Omega_t$ with $\lambda$ is identical to loading ground state atoms  with rate $\lambda$ into the system. The excitation avalanches of $\rho_{\vec{x},t}$ depend only on the difference $n_{\vec{x},t}-n_{c,t}=n_{\vec{x},t}+\lambda t -n_{c,0}$ and cannot distinguish between $n_{c,t}$ being decreased and $n_{\vec{x},t}$ being increased with rate $\lambda$. Experimentally, however, a controlled repopulation with rate $\lambda$ is often less feasible than adjusting the drive strength. 

In order to confirm the prediction of emergent SOC from the homogeneous treatment above and to observe its paradigmatic avalanche dynamics, we simulate the full time evolution of the Rydberg density via Eqs.~\eqref{Eq4} and \eqref{Eq5} in spatial dimensions $1\le d\le 3$. The equations are integrated on a $d$-dimensional grid of linear lattice spacing $\Delta x$ and we use dimensionless rates, expressed in units of $\Delta x^2/D$. The integration scheme is a derivative of the splitting scheme for stochastic differential equations with multiplicative noise \cite{Dornic2005}, adapted to the noise kernel of Eq.~\eqref{Eq4}, see Appendix~\ref{App1}.

For the simulations we set $\kappa_0=\Gamma=\frac{\Delta x^2}{2D}$,  $\tau_0=10^{-7}\Gamma$ and $\dr=10^{-2}\Gamma$, which is consistent with recent experiments~\cite{Buchhold2018a, Helmrich2018,Morsch2017}. Different degrees of scale separation are realized by varying $\lambda$ within the interval $\lambda\in [0,0.2\Gamma]$. We point out that, as for our choice of parameters, any realistic experiment will realize the conditions {\it (i)} and {\it (ii)} only on an approximate level. 

Our simulations reveal an extended dynamical regime, which is governed by the formation, propagation and decay of avalanches containing a significant number of excitations, $\rho_{\vec{x},t}\gg \frac{\lambda}{\dr}$, (see Fig.~\ref{Fig1}c). Parametrically it coincides well with the criterion $\tau_t<\lambda<\dr$, matching {\it (i)} and {\it (ii)}. In general, the distribution $P_{\text{ava}}(s)$ of avalanche sizes $s$ varies with $\lambda$. In the vicinity of a critical value $\lambda\approx\lambda_{soc}$ it, however, approaches a scale invariant form $P_{\text{ava}}(s)\sim s^{-\alpha}$ with an exponent $\alpha>0$.

In $d=1$, we obtain $\alpha= 1.44\pm0.1$, which is consistent with results obtained from other SOC models, e.g., the forest fire model \cite{Schenk} or activity patterns in the cortex \cite{Stewart}, and is associated with the underlying directed percolation universality class \cite{Hesse2014}. Its statistical error results from our sampling procedure, which dynamically counts avalanches from a finite number of patches of $10^4\times 10^4$ sites (time and space). For $d>1$, we predict $\alpha\approx1.5$, however, with larger errors due to our avalanche counting scheme. 

The scale invariant avalanche distribution is the hallmark of SOC \cite{Lu1995,Watkins2016,dickman2000}. It is accompanied by fractal spatio-temporal Rydberg excitation patterns (see Fig.~\ref{Fig1}c) and paradigmatic $\frac{1}{\omega}$-fluctuations \cite{Bak1987,Bak1988} in the Rydberg density $\rho_{\vec{x},\omega}\equiv\int \rho_{\vec{x},t}e^{i\omega t}dt\sim \omega^{-\beta}$, with $\beta \lessapprox1$ (see Fig.~\ref{Fig2}b). This clearly demonstrates a dynamical regime with SOC in the driven Rydberg gas. Its location at $\lambda\approx \lambda_{soc}$ can be understood as a trade-off in optimizing {\it (i)} and {\it (ii)} simultaneously for fixed values of $\tau_t, \dr$. For dimensions $d>1$ it approaches the estimate $\lambda_{soc}\sim\sqrt{\tau_t\dr}$. 

Moving $\lambda$ away from $\lambda_{soc}$, $P_{\text{ava}}(s)$ remains scale invariant in a finite range $|\lambda-\lambda_{soc}|<\eta$. We found $\eta\approx 0.2\lambda_{soc}$ for system sizes of $N=10^6$ lattice sites and our set of parameters. For larger deviations $|\lambda-\lambda_{soc}|>\eta$, the algebraic form of $P_{\text{ava}}(s)$ persists only for avalanche sizes $s<s_{||}(\lambda)$, i.e., below a $\lambda$-dependent cutoff scale $s_{||}(\lambda)$. Estimating the cutoff scale from the mean-field correlation length, i.e., $s_{||}(\lambda)=\xi_{||}$, which is justified far away from the SOC regime, one finds $s_{||}(\lambda)\sim \sqrt{\frac{D\dr}{2\kappa_t\lambda}}$ for $\lambda\gg\tau_t$ and $s_{||}(\lambda)\sim \sqrt{\frac{D\lambda}{\kappa_t\tau_t}}$ for $\lambda\ll\dr$. 

The behavior on distances above $s_{||}$ in the two regimes $\lambda\lessgtr\lambda_{soc}$ manifestly differs from each other. For supercritical values $\lambda\gg\lambda_{soc}$, the critical density $n_{c,t}$ decreases rapidly, leading to a large avalanche triggering rate and a high density of avalanches. On sizes $s>s_{||}(\lambda)$ different avalanches start to overlap, which makes them indistinguishable and generates a random excitation pattern (displayed in Fig.~\ref{Fig1}c), revealing the underlying avalanches only for $s<s_{||}(\lambda)$, (squares in Fig.~\ref{Fig1}d). 

The slow decrease of $n_{c,t}$ in the subcritical regime, $\lambda\ll\lambda_{soc}$ makes two subsequently following avalanches unfavorable and enforces a relative delay. It destroys the scale invariance above $s_{||}(\lambda)$ in favor of periodically triggered avalanches with increasing length $s\gg s_{||}(\lambda)$. This transforms the fractal real space structure found in the SOC regime into a time-periodic pattern, which is dominated by thermodynamically large excitation avalanches, shown in Fig.~\ref{Fig1}c. The period between two subsequent avalanches appears to be the time by which $n_{c,t}$ decreases by an integer value, i.e., $\delta t\approx \lambda^{-1}$. 

Our simulations reveal that the conditions {\it(i)} and {\it (ii)} do not have to be fulfilled exactly in order to realize avalanche dominated dynamics and self-organized criticality. We find SOC also for a broader parameter regime, which is approximately described by the condition 
\eq{rule}{
\tau_t\ll \lambda\ll\dr.
}
This condition can serve as a rule of thumb for the realization of self-organized criticality in experiments on driven Rydberg ensembles.

\section{Experimental observability} While the real space evolution of excitation avalanches is hard to access in experiments, the statistics of excitations, i.e., $\rho_{\vec{x},t}$ and $n_{\vec{x},t}$, can be measured via the particle loss rate $ \propto\dr \rho_{\vec{x},t}$~\cite{Helmrich2018, Buchhold2018a}. A robust, time-translational invariant observable is the integrated density \eq{Extra7}{R_t\equiv n_{0}+\lambda t-\int_{0}^t dt'\dr \langle\rho_{\vec{x},t'}\rangle_V,} where $n_0$ is the total initial density and $\langle ...\rangle_V=\frac{1}{V}\int_V d^dx$ denotes the spatial average over the system volume. Its meaning becomes clear when comparing it with the initial critical density $n_{c,0}$ at times $t\lambda\ll n_{c,0}$, yielding $R_t-n_{c,0}=\langle n_{\vec{x},t}\rangle_V-n_{c,t}$. 

Both $\rho_{\vec{x},\omega}$ and $R_t$ display very characteristic features in the three different regimes. For subcritical $\lambda$, the real time evolution of $R_t$ shows large, periodic amplitude fluctuations, reflecting individual, periodically triggered, extended avalanches. Instead, both the SOC and the supercritical regime feature much smaller amplitude fluctuations around $R_t\approx n_{c,o}$ (SOC) or $R_t\gg n_{c,o}$ (supercritical) as shown in Fig.~\ref{Fig2}a. In the subcritical (supercritical) regime, $\rho_{\vec{x},\omega}$ departs from its scale invariant form at SOC and one finds instead suppressed (pronounced) density fluctuations at intermediate frequencies, see Fig.~\ref{Fig2}b. 

Significant information is encoded in the statistics of $R_t$, especially its mean $\bar R \equiv \lambda \int_0^{\lambda^{-1}} R_t dt $ and fluctuations $\sigma_R^2\equiv\lambda \int_0^{\lambda^{-1}}R_t^2dt-\bar R^2$ as displayed in Fig.~\ref{Fig2}c. For subcritical $\lambda$ both $\bar R$ and $\sigma_R$ increase with $\lambda$ faster than the linear mean-field prediction. At the onset of SOC, however, both $\bar R$ and $\sigma_R$ experience a sharp drop, manifest in a non-analytic kink in their $\lambda$-dependence. While $\bar R\rightarrow n_{c,0}$ rapidly approaches the critical density, the fluctuations decrease by several orders of magnitude. Upon further increasing $\lambda$, $\bar R$ reaches a valley at $\approx n_{c,0}$ and subsequently increases again into the supercritical regime. $\sigma_R$ is featureless at the SOC-supercritical transition. 

\begin{figure}[t]\includegraphics[width=\linewidth]{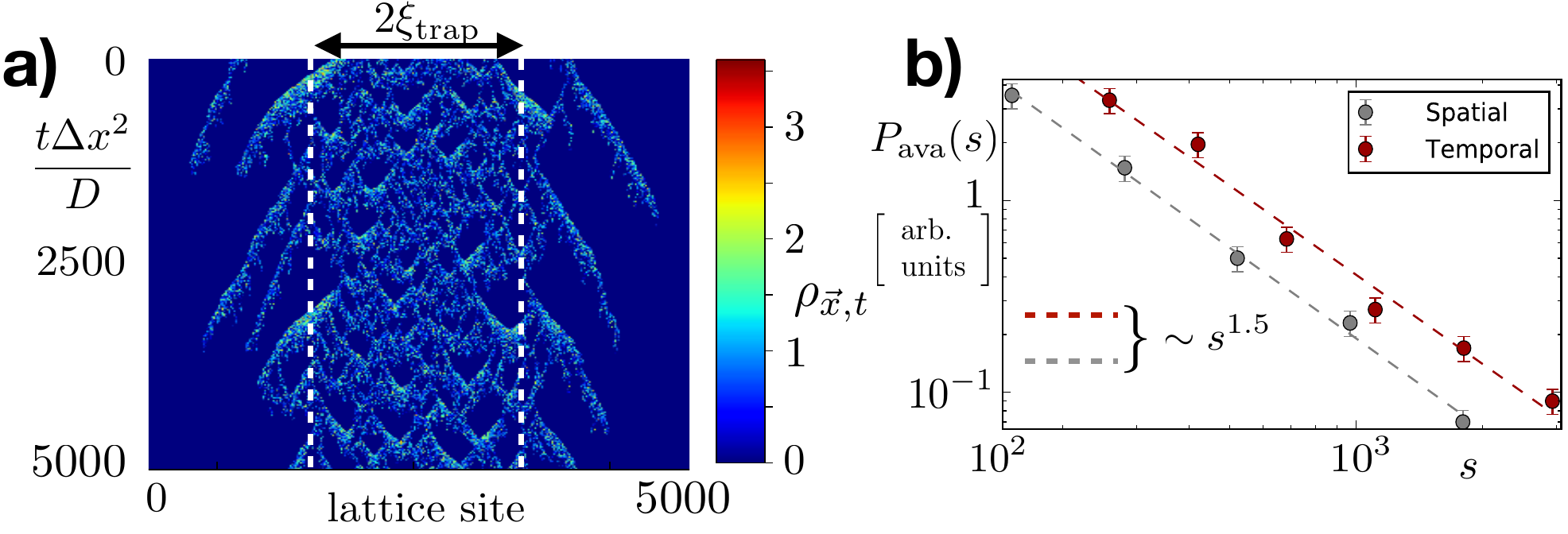}\caption{Avalanches in a trap ($d=1$).(a) Real space dynamics and(b) distribution of avalanches in a Gaussian trap of width $\xi_{\text{trap}}=10^3$ lattice sites in the SOC regime ($\lambda=2.36\times10^{-3}$). Both the spatial and the temporal avalanche size follow the same scaling exponent.}\label{Fig3}\end{figure}

In order to reason the observability of SOC for realistic conditions, where the atomic cloud is confined inside a trap, we expose $n_{\vec{x},t}$ to a potential of the form $V_{\text{trap}}(\vec{x})=V_0 \exp(-|\vec{x}|^2/\xi_{\text{trap}}^2)$, e.g., resulting from a Gaussian trapping laser with beam waist $\xi_{\text{trap}}$ \cite{Buchhold2018a}. For a mean free path $d_a\ll\xi_{\text{trap}}$, the effect of $V_{\text{trap}}(\vec{x})$ can be treated within the relaxation time approximation, see Appendix~\ref{App2}. This adds a drift $\sim-\vec{v}_{\vec{x}}\cdot \vec{\nabla}n_{\vec{x},t}$ to the right-hand side of Eq.~\eqref{Eq5}. Here $\vec{v}_{\vec{x}}=\frac{d_a}{\sqrt{Mk_{\text{B}}T}}\vec{\nabla}V_{\text{trap}}$ is the relaxation velocity. The dynamics following this drift at low temperatures $T$ ($\frac{V_0d_a}{\sqrt{Mk_{\text{B}}T}}=0.7D$) is displayed in Fig.~\ref{Fig3}a. On distances $|\vec{x}|<\xi_{\text{trap}}$, avalanches remain well defined and both their fractal real space pattern and the scale invariant statistics are observable below the trap scale, see Fig.~\ref{Fig3}b.

\section{Effect of the spatial dimension} Apart from Rydberg atoms, the continuum model in Eq.~\eqref{Eq4} may also serve as a coarse grained description for activity spreading in sparse networks \cite{Hesse2014}. In this picture, each Rydberg atom represents a node and the parameters $\kappa, \tau, \Gamma$ describe its reaction to external stimuli and the decay of information. The density $n_{\vec{x},t}$ represents a 'node energy', which is consumed by active nodes with rate $\dr\rho_{\vec{x},t}$ and recharged with rate $\lambda$.

Optimal networks are expected to operate close to SOC \cite{Arca2006,Shew2015,Hesse2014,Markovic2014,Kinouchi2006,Levina2007}. Their natural tuning parameter is the average connectivity $z$ of the nodes, which is adjusted to match external conditions \cite{Levina2009, Bornholdt2000, Bertsch2004, Kinouchi2006, Levina2007}. Figure~\ref{Fig2}c confirms that here the dimensionality $d$ acts as a second 'control parameter'. Changing $d$ from $d=1$ to $d=2$ shifts the scale invariant regime (shaded region) and increases its range. For a given set $\tau, \lambda, \dr$, there may exist an 'optimal' $d$, for the system to display SOC. In Rydberg experiments $d$ can be controlled by adjusting the trapping geometry. Combined with the tuneability  of $\lambda$ and $\tau$, this offers many possibilities to study self-organized criticality in network-like setups.

\section{Conclusion}
We propose and study an experimentally feasible mechanism to control excitation avalanches in driven Rydberg setups~\cite{Buchhold2018a}. On large, transient times, one can observe subcritical, supercritical and self-organized critical avalanche dynamics, depending on the control parameter. Each regime features unique signatures, including a scale invariant avalanche distribution and $\frac{1}{\omega}$-noise, both paradigmatic signals for SOC. This motivates driven Rydberg ensembles~\cite{Buchhold2018a} as viable platforms for the study of SOC and the conditions under which simple dynamical rules, as imposed by the facilitation condition, can establish and maintain self-ordering towards complex dynamics structures. 

While the crossover from the SOC to the supercritical regime does not produce a pronounced feature in the integrated density, Fig.~\ref{Fig2} reveals a developing non-analyticity in both the integrated density as well as its fluctuations as $\tau_t$ is decreased. It hints towards an underlying critical point, on the one hand such a critical point might describe the SOC universality class, including avalanche and correlation exponents. On the other hand, it could be a remnant of the directed percolation critical point, which would be reached for $\lambda,\tau\rightarrow0$. In both cases, the investigation of this conjectured critical point and its relation to the SOC universality seems worthwhile for future work.

Based on the similarity of the corresponding master equations, we conjecture a relation between driven Rydberg gases and self-organizing neural networks. The analogy is strengthened by frequently observed periodic or random activity patterns in non-optimal operating networks~\cite{Prinz5953,Ong2012}. Exploring this connection, especially for the role that is played by scale separation, appears a promising direction to connect driven Rydberg systems with neurosciences. 

\acknowledgements We thank G.~Refael, S.~Diehl and S.~Whitlock for valuable comments on the manuscript. K.~K. was supported by the J. Weldon Green SURF fellowship and M.~B. acknowledges support from the Alexander von Humboldt foundation.

\appendix
\section{Numerical integration scheme}\label{App1}
Numerical integration of Eqs. \eqref{Eq4} and \eqref{Eq5} is performed by an operator-splitting update scheme \cite{Dornic2005}. At each time step, the evolution is decomposed into a stochastic evolution step and a deterministic step. The former is designed to solve a stochastic differential equation of the form:
\eq{StochasticStep}{
\partial_t \rho_{\vec{x},t} &=& \alpha + \beta\rho_{\vec{x},t} + \sigma\sqrt{\rho_{\vec{x},t}}\eta.
}
Here $\eta$ is a Markovian noise kernel with mean zero and unit variance. For small $\gamma_{\downarrow 0}, \kappa, \tau$, we may approximate $\alpha$ and $\beta$ to be constant over each time step.  The corresponding Fokker-Planck equation has the exact solution 
\eq{FokkerPlanck}{
P(\rho, \delta t) = \lambda e^{-\lambda\left(\rho_0e^{\beta \delta t} + \rho \right)}\left(\frac{\rho}{\rho_0e^{\beta \delta t}} \right)^{\mu/2}I_{\mu}\left(2\lambda\sqrt{\rho_0\rho e^{\beta \delta t}} \right),
}
where we set $\rho\equiv\rho_{\vec{x},t+\delta}$ and $\rho_0\equiv\rho_{\vec{x},t}$ as well as $\lambda = \frac{2\beta}{\sigma^2(e^{\beta t} - 1)}$ and $\mu = \frac{2\alpha}{\sigma^2} - 1$ and $I_{\mu}(x)$ is the modified Bessel function of the first kind with index $\mu$ and argument $x$. This can be expressed via a mixed Gamma distribution which allows for efficient sampling:
\eq{MixedGamma}{\rho=\Gamma[\mu + 1 + \text{Poisson}[\lambda\rho_0 e^{\beta \delta t}]]/\lambda,}
which is shorthand notation for a random variable which is drawn from a Gamma distribution with argument $\mu+1+x$, whereas $x$ was drawn from a Poisson distribution with argument $\lambda\rho_0 e^{\beta\delta t}$. 

Given the values of $\rho_{\vec{x},t}$ at time $t$, its stochastic evolution $\rho_{\vec{x},t+\delta t}$ after a step $\delta t$ can be drawn from the above distribution. The deterministic part of the equation of motion has a purely polynomial form and can also be solved exactly. The time discretization error is therefore only caused by the splitting of the evolution into a stochastic and a deterministic part.  

A non-zero $\tau_t$ can be incorporated by using the same procedure with a simple change of variables: $u = \rho + \tau_t/\kappa_t$. The non-negativity of $\rho$ is enforced after sampling by resetting any value of $u<\tau_t$ to $\tau$. The well-behaving evolution of $n_{\vec{x},t}$ is performed via an Euler scheme.

\section{ Relaxation time approximation in a trap}\label{App2} In the presence of an inhomogeneous background potential $V(\vec{r})$ for the particles, the drift term in Eq.~\eqref{Eq7} becomes significant. For the active density it yields \eq{e123}{\partial_t n_{\vec{x},t}&=&\vec{\nabla}\sum_{l}\Theta(r_{\text{fac}}-|\vec{x}-\vec{r}_l(t)|)\langle \sigma^{rr}_l+\sigma^{gg}_l\rangle_t\frac{\vec{p}_l}{M}\\ &&+\sum_{l}\Theta(r_{\text{fac}}-|\vec{x}-\vec{r}_l(t)|)\partial_t\langle \sigma^{rr}_l+\sigma^{gg}_l\rangle_t\nonumber} where we applied the chain rule and inserted the momentum $\vec{p}_l=M\partial_t\vec{r}_l$. In the relaxation time approximation, the momentum $\vec{p}$ is reset after a characteristic scattering time $t_{\text{rel}}=d_a{\sqrt{\frac{M}{2 \pi k_{\text{B}}T}}}$, where $d_a$ is the mean free path and $T$ is the temperature.  This yields the equation of motion \eq{e124}{\partial_t\vec{p_l}=-\vec{\nabla}V(\vec{r}_l)-\frac{1}{t_{\text{rel}}}\vec{p}_l.} It is stationary for $\vec{p}_l=-t_{\text{rel}}\vec{\nabla}V(\vec{r}_l)$ and induces an average drift for times $t>t_{\text{rel}}$. Inserting this result in Eq.~\eqref{e123} and neglecting the variation of $V$ on length scales $\sim r_{\text{fac}}$, i.e., $V(\vec{r}_l)\approx V(\vec{x})$, one finds \eq{e125}{\partial_t n_{\vec{x},t}&=&-\frac{d_a}{\sqrt{2Mk_{\text{B}}T}}\vec{\nabla}V(\vec{x})\vec{\nabla}n_{\vec{x},t}+...,}where $...$ describes the dynamics of the internal states of the atoms. This approximation works well if both the facilitation shell and the mean free path are much smaller than the typical length scale of the potential $V$. 

\bibliography{soc}

\end{document}